\documentstyle[12pt,epsf]{article}
\begin{document}

\begin{titlepage}

\setcounter{page}{1}
\rightline{JLAB-THY-98-07}
\rightline{WM-98-102}
\vfill
\begin{center}
 {\Large \bf QCD Evolution Equations: } 
{\large \bf Numerical Algorithms from the Laguerre Expansion}

\vfill
\vfill
 {\large 
Claudio Corian\`{o}$^{**}$\footnote{
        E-mail address: coriano@jlabs2.jlab.org} and    
        Cetin Savkli$^{**\dagger}$\footnote{csavkli@physics.wm.edu}
}

\vspace{.12in}
 {\it $^{**}$ Theory Group, Jefferson Lab, Newport News, VA 23606, USA}

\vspace{.12in}
 {\it $^{\dagger}$ Dept. of Physics, College of William and Mary, 
Williamsburg, VA, 23187 }
\vspace{.075in}

\end{center}
\vfill
\begin{abstract}
A complete numerical implementation, in both singlet and non-singlet sectors, 
of a very elegant method to solve the 
QCD Evolution equations, due to Furmanski and Petronzio, is presented. 
The algorithm is directly implemented in x-space by a Laguerre expansion 
of the parton distributions.
All the leading-twist distributions are evolved: 
longitudinally polarized, transversely polarized and unpolarized, 
to NLO accuracy. The expansion is optimal at finite x, up to reasonably small x-values ($x\approx 10^{-3}$), below which the convergence of the expansion slows down.  
The polarized evolution is smoother, due to the 
less singular structure of the polarized DGLAP kernels at small-x.   
In the region of fast convergence, which covers most of the usual 
perturbative applications, high numerical accuracy is achieved 
by expanding over a set of approximately 30 polynomials, 
with a very modest running time.

\end{abstract}
\smallskip

\end{titlepage}

\setcounter{footnote}{0}

\def\beq{\begin{equation}}
\def\eeq{\end{equation}}
\def\beqn{\begin{eqnarray}}
\def\eeqn{\end{eqnarray}}
\def\ie{{\it i.e.}}
\def\eg{{\it e.g.}}
\def\half{{\textstyle{1\over 2}}}
\def\nicefrac#1#2{\hbox{${#1\over #2}$}}
\def\third{{\textstyle {1\over3}}}
\def\quarter{{\textstyle {1\over4}}}
\def\m{{\tt -}}

\def\p{{\tt +}}

\def\slash#1{#1\hskip-6pt/\hskip6pt}
\def\slk{\slash{k}}
\def\GeV{\,{\rm GeV}}
\def\TeV{\,{\rm TeV}}
\def\y{\,{\rm y}}

\def\l{\langle}
\def\r{\rangle}

\setcounter{footnote}{0}
\newcommand{\beqa}{\begin{eqnarray}}
\newcommand{\eeqa}{\end{eqnarray}}
\newcommand{\eps}{\epsilon}

\pagestyle{plain}
\setcounter{page}{1}
\section*{{Program Summary}}
{\em Title of the Programs:} nsunpol, snunpol, nslong, snlong, trans\\

\noindent
{\em Computer:} \\ Sun 19\\

\noindent
{\em Operating system:} Unix\\

\noindent
{\em Programming language used:} FORTRAN 77\\

\noindent
{\em Peripherals used:} Laser Printer\\

\noindent
{\em Number of lines in distributed program:} 
4260\\

\noindent
{\em Keywords:} Structure function, polarized parton distribution, $Q^2$ 
evolution, Laguerre expansions, numerical solution.\\

\noindent
{\em Nature of physical problem:}\\
The programs provided here solve the DGLAP evolution equations, with 
next-to-leading order $\alpha_s$ effects taken to account, 
for unpolarized, longitudinally 
polarized and transversely polarized parton distributions.  \\

\noindent
{\em Method of solution:}\\
The method developed by Furmanski and Petronzio is used. The kernel $P(x)$ of 
the DGLAP integrodifferential equations and the 
evolution operators $E(t,x)$ are 
expanded in Laguerre polynomials.


\noindent
{\em Typical running time:}\\
About 5 seconds for the transverse polarization case and 30 minutes 
for the longitudinal polarization and for the unpolarized.

\section*{{\bf LONG WRITE-UP}}
  
\section{Introduction}
The study of the spin structure of the proton is a fascinating 
aspect of the theory of the strong interactions. 
Parton distributions in the nucleon tell us about the structure of fundamental observables such as spin, parton densities 
and correlations among partons, in a light-cone framework. In 
a second quantized theory they emerge as matrix elements 
of nonlocal operators at light-like separations. At low energy they can be described by 
valence quark models and the impact of the QCD evolution (or Renormalization Group Evolution, RGE) 
on the shape of these distributions can be performed within the parton model. 

It has become more and more common in the high energy literature to describe 
the evolution of parton distributions by starting from a low energy input. 
 
There is widespread interest in the analysis 
of the effect of the QCD evolution especially for those 
leading twist distributions (such as the transverse spin distribution) 
which have not yet been measured. For instance, in ref.~\cite{vento}, the authors 
analize the impact of the evolution 
on transverse spin distributions derived at low $Q$ 
from the Isgur-Karl quark model \cite{IK}, and these predictions can be used direclty -at very high energy- for other predictions, 
such as in the study of the transverse spin dependence of the Drell Yan process. 
A similar analysis can be done for the chiral quark model \cite{HW}.

On the numerical side, various methods have been presented in the 
literature \cite{Kumano} 
which all try to solve the DGLAP equations by iterating the evolution over 
infinitesimal steps in the fractional momentum $x$. 
Another technique is based on the use of Mellin moments and on their inversion.In general, moments are equivalent to finite -rather than infinitesimal- 
discretizations of the integro-differential equations. 
In the case of the Mellin moments the inversion 
(to x-space) is the real difficult and time consuming part of the method. 

A different, and very elegant implementation of methods based on finite discretizations 
of RGE's for QCD was formulated long ago by Furmanski and Petronzio 
\cite{FP1}. 
Their method uses the Laguerre expansion of the 
initial distributions and of the kernel of the evolution equations arrested to an 
arbitrary order $n$. The algorithm that they provide defines the structure of the moments 
recursively in terms of some initial conditions. In the non-singlet sector there have 
been attempts to apply the method both to leading and to next to leading order 
\cite{Ramsey,Kumano1}, and to leading order in the singlet sector as well \cite{Ramsey}. 
However, a complete implementation and numerical documentation of this algorithm 
is still missing. 
We also mention that our interest in this method has aroused 
from our search for the numerical implementations of more involved evolutions equations, 
such as those describing the dynamics of Compton scattering 
in the deeply virtual limit (DVCS) \cite{JR}.
In this latter case, the RGE evolution is a continuous interpolation between 2 limits: the DGLAP (or Altarelli Parisi)
evolution and the Efremov-Radyushkin-Brodsky-Lepage evolution (ERBL) \cite{ERBL}. 
We believe that a complete numerical understanding  of these ``non-diagonal''  RGE's and their 
robust numerical implementation requires finite step integrations.  
We hope to get back to this issue in the near future, and for the rest of this paper 
just focus on the usual DGLAP evolution.    
Here we have implemented the evolution of all the leading twist 
parton distributions, 
using the kernels calculated by various authors \cite{FP1,FP2,MVN,Vogel}. 
A complete list of these kernels can be found in the Appendix.

\section{The Laguerre expansion}

The Laguerre method for the numerical solution of the evolution equations is due to Furmanski and Petronzio. In this section we briefly outline 
the method, which is fast converging at intermediate values of x 
and any $Q^2$ value. At very small-x (approximately $10^{-3}$), 
the Laguerre expansion suffers from numerical instabilities, due to the growth of 
the moments.
We start by defining our notation and other conventions. 

The two-loop running of the coupling constant is defined by 

\beq
{\alpha(Q_0^2)\over 2 \pi}={2\over \beta_0}
{ 1\over \ln (Q^2/\Lambda^2)}\left( 1 - {\beta_1\over \beta_0}
{\ln\ln(Q^2/\Lambda^2)\over \ln(Q^2/ \Lambda^2)} +
O({1\over\ln^2(Q^2/\Lambda^2) })\right).
\eeq

where

\beqa
&& \beta_0={11\over 3} C_G - {4\over 3}T_R n_f \nonumber \\
&& \beta_1={34\over 3} C_G^2 - {10\over 3} C_G n_f - 2 C_F n_f
\eeqa
where 

\beq
C_G= N, \,\,\,\,\,\,\, C_F={ N^2-1\over 2 N},\,\,\,\,\,\,\,\, T_R={1\over 2}
\eeq

and N is the number of colours. 

The solution for the running coupling is given by 
\beqa
\alpha(t)&=&{\alpha(0)\over 2 \pi} e^{-\beta_0/2 t}
\eeqa
with $\alpha(Q_0^2)\equiv \alpha(0) $, and $Q_0$ denoting the 
initial scale at which the evolution starts. 
The evolution equations are of the form 

\beqa  
 Q^2 {d\over d Q^2} {q_i}^{(-)}(x, Q^2) &=& 
{\alpha(Q^2)\over 2 \pi} P_{(-)}(x, \alpha(Q^2))\otimes q_{i}^{(-)}(x, Q^2)
\nonumber \\
 Q^2 {d\over d Q^2}\chi_i(x,Q^2)
&=& 
{\alpha(Q^2)\over 2 \pi} P_{(-)}(x, \alpha(Q^2))\otimes \chi_i(x,Q^2)
\nonumber \\
\eeqa
with 
\beq
\chi_i(x,Q^2)={q_i}^{(+)}(x, Q^2) -{1\over n_F} 
q^{(+)}(x, Q^2)
\eeq
for the non-singlet distributions and 
\beqa
&& Q^2 { d\over d Q^2} \left( \begin{array}{c}
q^{(+)}(x, Q^2) \\
G(x, Q^2) \end{array}\right)=
\left(\begin{array}{cc}
P_{qq}(x,Q^2) & P_{qg}(x,Q^2) \\
P_{gq}(x,Q^2) & P_{gg}(x,Q^2)
\end{array} \right)\otimes 
\left( \begin{array}{c}
q^{(+)}(x, Q^2) \\
G(x,Q^2) \end{array}\right) \nonumber \\
\eeqa
for the singlet sector. 

We have defined, as usual 

\beq
q_i^{(-)}=q_i -\bar{q}_i, \,\,\,\,\,
q^{(+)}_i= q_i + \bar{q}_i, \,\,\,\,\,\,
q^{(+)}\equiv\Sigma= \sum_{i=1}^{n_f} q^{(+)}_i.
\eeq
We introduce the evolution variable
\beq
t= -{2\over \beta_0}\ln {\alpha(Q^2)\over \alpha(Q_0^2)}
\eeq
which replaces $Q^2$. 
The evolution equations are then rewritten in the form

\beqa
 {d\over dt} q_i^{(-)}(t,x)&=&\left( P^{(0)}(x) + 
{\alpha (t)\over 2 \pi}R_{(-)}(x) +...\right)\otimes q_i^{(-)}(t,x)
\label{one} \\
 Q^2 {d\over d t}\chi_i(x, Q^2)
&=& 
\left( P^{(0)}(x) +
{\alpha(t)\over 2 \pi}R_{(+)}(x)\right)\otimes
\chi_i(x,Q^2),
\label{two} \\
{d\over d t}\left( \begin{array}{c}q^{(+)}(t,x) \\
G(x,t)
\end{array} \right)&=& \left( P^{(0)}(x) +
{\alpha(t)\over 2 \pi} R(x) +...\right)\otimes 
\left( \begin{array}{c}
q^{(+)}(x,t) \\
G(x,t) 
\label{three}
\end{array} \right).
\eeqa
In the new variable $t$, the kernels of the evolution take the form 

\beqa
 R_{(\pm)}(x)& = &P^{(1)}_{(\pm)}(x) - 
{\beta_1\over 2 \beta_0} P^{(0)}_V(x)
\nonumber \\
 R(x) &=& P^{(1)}(x)-{\beta_1\over 2 \beta_0} P^{(0)}(x).
\eeqa
Equations (\ref{one}) and (\ref{two}) are solved 
independently for the variables $q_i^{(-)}$ and 
$\chi_i$ respectively. Finally, the solution $q^{(+)}$ 
of eq.~(\ref{three}) (or the singlet equation) 
is substitued into $\chi_i$ in order to obtain $q_i^{(+)}$. 
The equations can be written down in terms of two singlet evolution operators 
$E_{\pm}(t,x)$ and initial conditions 
$\tilde{q}_{\pm}(x,t=0)\equiv \tilde{q}_{\pm}(x)$ as 

\beq
{d\over dt} E_{\pm}=P_{\pm}\otimes E_{\pm},
\eeq
whose solutions are given by
\beqa
&& q_i^{(-)}(t,x)= E_{(-)}\otimes \tilde{q}_i^{(-)} \nonumber \\
&& \chi_i(t,x)=E_{(+)}\otimes \tilde\chi_i(x).
\eeqa
The singlet evolution for the matrix operator $E(t,x)$

\beqa
&& \left( \begin{array}{cc}
E_{FF} & E_{FG} \\
E_{GF} & E_{GG} 
\end{array} \right)
\eeqa

\beq
{d E\over d t}= P\otimes E
\eeq

is solved similarly as 

\beqa
&&\left( \begin{array}{c}
 q^{(+)}
(t,x)\\
G(t,x)
\end{array}\right)= E(t,x)\otimes 
\left( \begin{array}{c}
\tilde{q}^{(+)}(x)\\
\tilde{G}(x)
\end{array}\right). \\
\eeqa

The method of Furmanski and Petronzio requires 
an expansion of the splitting functions and of the parton distributions in the 
basis of the Laguerre polynomials 

\beq
L_n(y)=\sum_{k=0}^n \left(\begin{array}{c} n \\ k \end{array} \right)
(-1)^k {y^k\over k!}
\eeq

which satisfies the property of closure under a convolution

\beq
L_n(y)\otimes L_m(y)=L_{n+m}(y) - L_{n +m +1}(y).
\eeq
In order to improve the small-x behaviour of the algorithm, from now on, 
the evolution is applied to the modified kernel $x P(x)$, which, 
for simplicity, is still denoted as in all the equations above, 
i.e. by $P(x)$. At a second step, the $ 0<x<1$ interval 
is mapped into an infinite interval $0<y<\infty$ by a change of variable 
$x=e^{-y}$ and all the integrations are performed in this last interval.  
We start from the non-singlet case by defining 
the Laguerre expansion of the kernels and the corresponding (Laguerre) moments to lowest order 

\beqa
 P_V^{(0)}(y)&=&\sum_{n=0}^\infty P_n^{(0)} L_n(y), \nonumber \\
P_n^{(0)}&=&\int_0^\infty dy e^{-y} L_n(y)\,\, P^{(0)}(y)
\eeqa

and to NLO

\beq
R(y)=\sum_{n=0}^\infty R_n L_n(y).
\eeq
One defines also the difference of  moments 

\beqa
 p_i^{(0)}&=&P_i^{(0)} -P_{i-1}^{(0)}\,\,\,\,\,\,\,(P_{-1}^{(0)}=0) \nonumber \\
r_i&=& R_i- R_{i-1} \,\,\,\,\,\, R_{-1}=0.
\eeqa

A similar expansion is set up for the evolution operators  $E(t,y)$

\beqa
E^{(0)}(t,y)&=&\sum_{n=0}^\infty E_n^{(0)}(t) L_n(y)\nonumber \\
E(t,y)&=&\sum_{n=0}^\infty E_n(t) L_n(y),
\eeqa

where all the information on the $t$ evolution is contained in the moments $E_n(t)$. The solution to NLO is expressed as \cite{comm}

\beqa
 E_n (t)=E_n^{(0)}(t) -{2\over \beta_0}{\alpha(t)-\alpha(0)\over 2\pi}E_n^{(1)}(t),
\label{solve}
\eeqa

where
\beqa
&& E_n^{(0)}(t)=e^{P_0^{(0} t} \sum_{k=0}^n {A_n^{(k)} t^k\over k!} \\
&& E_n^{(1)}(t)=\sum_i^n r_{n-i} E_i^{(0)}(t), \\
\eeqa
and the coefficients $A_n^{(k)}$ are determined recursively from the moments of the  
lowest order kernel $P^{(0)}$

\beqa
 A_n^{(0)}&=& 1 \nonumber \\
 A_n^{(k+1)}&=&\sum_{i=k}^{n-1}p_{n-i}^{(0)} A_i^{(k)} \hspace{1cm} (k=0,1,2,...,n-1).
\eeqa

In the singlet case one proceeds in a similar way. The solution is expressed in terms of a 
2-by-2 matrix operator

\beq
E^{(0)}(t,y)=\sum_{n=0}^\infty E_n^{(0)}(t) L_n(y).
\eeq
The solution (at leading order) is written down in terms of 2 projection matrices and one 
eigenvalue ($\lambda$) of the $P^{(0)}$ (matrix) kernel

\beqa
e_1={1\over \lambda}P^{(0)}, \hspace{2cm} 
e_2={1\over \lambda}\left( -P^{(0)} +\lambda \bf{1}\right), \\
\eeqa
where
\beqa
\lambda&=&-({4\over 3} C_F +{2\over 3} n_f T_R),
\eeqa

in the form
\beqa
E^{(0)}_n(t)&=&\sum_{k=0}^n {t^k\over k!}\left( A_n^{(k)}+ B_n^{(k)}e^{\lambda t}\right).
\eeqa

The recursion relations which allow to build $A_n^{(k)}$ and $B_n^{(k)}$ are solved in two 
steps as follows. One solves first for two sets of matrices $a_n^{(k)}$ and $b_n^{(k)}$  by the relations

\beqa
a_n^{(0)}&=&0 \nonumber \\
 a_n^{(k+1)}&=&\lambda e_1 a_n^{(k)}+\sum_{i=k}^{n-1}p_{n-i}^{(0)} a_i^{(k)}\nonumber \\
 b_n^{(0)}&=& 0 \nonumber \\
 b_n^{(k+1)}&=& -\lambda e_2 b_n^{(k)} +\sum_{i=1}^{n-1} p_{n-i}^{(0)} b_i^{(k)},
\label{solve1}
\eeqa

which are used to construct the matrices $A_n^{(0)}$ and $B_n^{(0)}$

\beqa
A_n^{(0)}&=& e_2 -{1\over \lambda^n}\left( e_1 a_n^{(n)} - (-1)^n e_2 b_n^{(n)}\right)\nonumber \\
B_n^{(0)}&=& e_1 +{1\over \lambda^n}\left( e_1 a_n^{(n)} - (-1)^n e_2 b_n^{(n)}\right) . \\
\eeqa

These matrices are then input in the recursion relations 

\beqa
A_0^{(0)}&=& e_2 \hspace{1cm}B_0^{(0)}= e_1 \nonumber \\
A_n^{(k+1)}&=& \lambda e_1 A_n^{(k)} +\sum_{i=k}^{n-1}p_{n-i}^{(0)} A_i^k \nonumber \\
B_n^{k+1}&=& -\lambda e_2 B_n^{(k)}+\sum_{i=k}^{n-1} p_{n-i}^{(0)} B_i^{(k)}
\eeqa
witn $n>0$ and $k=0,1,...,n-1$, 
which generates the coefficients of the matrix-valued operator $E^{(0)}$ 
(i.e. the leading order solution). The NLO part of the evolution is obtained from 

\beqa
E^{(1)}(t,y)&=&\sum_{n=0}^{\infty} E_n^{(1)}(t) L_n(y),
\eeqa

with
\beqa
E_n^{(1)}(t)&=&\tilde{E}_{n}^{(1)}(t)-2 \tilde{E}_{n-1}^{(1)}(t)+ \tilde{E}_{n-2}^{(1)}(t)
\eeqa

where

\beq
\tilde{E}_{n}^{(1)}(t)=\int_o^t d\tau e^{-\beta_0 \tau/2}\sum_{ijk} E^{(0)}(t-\tau)
R_j E_k^{(0)}(\tau) \delta(n-i-j-k).
\eeq
The expressions of $E^{(0)}$ and $E^{(1)}$ are inserted into eq.~(\ref{solve}) thereby 
providing a complete NLO solution of the singlet sector. 

\section{ The polarized and the unpolarized evolution}

The implementation of the polarized and of the unpolarized evolution is performed in the $\overline{MS}$ scheme, which is by now standard in most of the high energy physics applications. 
In the unpolarized case, we introduce 
valence quark distributions $q_V(x,Q_0^2)$ and gluon distributions $G(x,Q_0^2)$ at the input scale 
$Q_0$, taken from the 
CTEQ parametrization \cite{cteq}
\beqn
q(x)&=&A_0 x^{A_1}(1-x)^{A_2}(1 + A_3 x^{A_4}).
\eeqn

Specifically

\beqn
x u_V(x)&=&1.344x^{0.501}(1-x)^{3.689}[1 + 6.042 x^{0.873}] \nonumber \\
x d_V(x)&=&0.640 x^{0.501}(1-x)^{4.247}[1 + 2.690 x^{0.333}] \nonumber \\
x G(x)&=&1.123 x^{-0.206}(1-x)^{4.673}[1 + 4.269 x^{1.508}]
\eeqa

and an asymmetric sea contribution
\beqn
x\overline{q}(x)&=&{1\over 2}[0.255 x^{-0.143}(1-x)^{8.041}(1 +6.112 x)\mp 0.071 x^{0.501}(1-x)^{8.041}
],
\eeqn

where the $(-)$ holds for the $\bar{u}$ and the $(+)$ for the $\bar{d}$ flavors. The set accounts for a 
$\bar{u}, \bar{d}$ flavour asymmetry, and the sea quark contribution is parameterized by 

\beqn
x \bar{s}(x)&=&[0.064 x^{-0.143}(1-x)^{8.041}(1+ 6.112 x)]. 
\eeqn

In the polarized case we have chosen the first set of ref.~\cite{GS} which is of the functional form 
$A B x^C(1-x)^D(1 + E x + F \sqrt{x})$

\beqn
x\Delta u_V&=&0.918* 1.365* x^{0.512}(1-x)^{3.96}(1 + 11.65 x- 4.6 \sqrt{x})
\nonumber \\
x\Delta d_V&=&-0.339 *3.849 x^{0.78}(1-x)^{4.96}(1 + 7.81 x - 3.48 \sqrt{x})
\nonumber \\
x\Delta G&=&1.71 *3.099 x^{0.724}(1-x)^{5.71}(1 + 0.0 x + 0.0 \sqrt{x}).
\eeqn

For the (flavour symmetric) sea contribution we have set $\Delta\bar{u}=\Delta\bar{d}=\Delta\bar{s}$
with 
\beq
\Delta\bar{s}=-0.06* 18.521 x^{0.724}(1-x)^{14.4}(1+ 4.63 x - 4.96\sqrt{x}).
\eeq

Notice that the evolution in the $\overline{MS}$ requires some care, depending upon the way 
the subtraction of the collinear singularities in the coefficient functions (hard scatterings) is performed. The (non singlet) hard scatterings, in fact, do not conserve helicities in the annihilation channels. In a ``traditional'' 
$\overline{MS}$ scheme, one has to keep both these helicity violating 
terms of the coefficient functions 
and has add to the polarized non-singlet kernels of the Appendix some additional terms proportional to $(1-x)$ to NLO 
\cite{CCFG,vogel1,MVN} 
As a result of this, both the singlet and the non-singlet evolution are affected to NLO. However, one can factorize out of the coefficient functions these spuriuos (helicity-violating) 
terms and absorb them into the renormalization of the parton distributions. As a result of this procedure, the NLO 
kernels of the evolution turn out to be exactly those defined in the Appendix and 
the hard-scatterings are helicity preserving.

\section{The evolution of the transverse spin distribution}

The first identification 
of the transverse spin distribution is due to Ralston and Soper \cite{RS} 
in their study of the factorization formula for the Drell Yan cross section.  
The parton interpretation of this distribution has been 
discussed in various papers \cite{artru,JJ}, in which its 
behaviour as a leading twist distribution has been pointed out.

It appears in the double transverse spin asymmetry $A_{TT}$ 
for the Drell Yan process \cite{RS,JJ}

\beq
A_{TT}={\sin^2\theta \cos 2\phi\over 1 + \cos^2\theta}
{\sum_ie_i^2\Delta_T q_i(x_1)\Delta_T \bar{q}_i(x_2)\over 
\sum_ie_i^2 q_i(x_1) \bar{q}_i(x_2)}
\label{ralso}.
\eeq

In eq.~(\ref{ralso}) the angles $\theta$ and $\phi$ are the polar and the 
azimuthal angles of the momentum of one of the two leptons, 
measured with respect to the beam 
($\theta$) and to the photon polarization directions ($\phi$). The 
asymmetry disappears if the momenta of the two leptons 
are both integrated over.   
In the parton model, $\Delta_T(q)$ is interpreted as the probability 
of finding a quark with spin polarized along the transverse spin of a transversely polarized proton minus the probability to find it polarized oppositely. 
This distribution does not couple to gluons and is therefore, 
purely non-singlet. The LO anomalous dimensions 
of this distribution have been calculated in \cite{artru}, while 
the NLO corrections have been derived by various authors \cite{vogel1}, 
using both the Operator Product Expansion and the method 
of ref.~\cite{FP3}, extended to the polarized case. 

Since the gluons don't couple in the evolution, 
the equation is simply written as

\beq
{\partial\over \partial \ln Q^2} 
\Delta_T q_{\pm}(x,Q^2)={\alpha_s(Q^2)\over 2 \pi} 
\Delta_T P_{q^{\pm}} (x)\otimes \Delta_T q_{\pm}(x,Q^2).
\eeq

As in the previous sections, we have set 
$\Delta_T q_{\pm}=\Delta_T q \pm\Delta_T \bar{q}$.

\section{Description of the program}
The first step in the calculation is the computation of Laguerre 
moments -or expansion coefficients- of the kernel which are given by an explicit 
recursion relation.

The solutions of the integral equations are obtained by first discretizing the 
integrals in the form 
\begin{equation}
\int dq\, f(q) \longrightarrow \sum_{i=1}^n w_i\, f(q_i),
\end{equation}
where $w_i$ are integration weights for the grid-points $q_i$. Various sets of Gauss-Legendre 
grid-points are provided (by the GAUSS and the LEGENDRE subroutines) in the interval $(-1,1)$ . 
In order to map 
the grid points and the weights from the interval $(-1,1)$ to the interval $(0,\infty)$, 
one can use various mappings. The possible types of mapping provided in the code are \\
\noindent
(i) MAPNO=1: 
\begin{equation}
y(x)=R_{min}+\frac{(1+x)}{1-x+2/(R_{max}-R_{min})},
\label{mapno1}
\end{equation}
(ii) MAPNO=2: 
\begin{equation}
y(x)=R_{min}+(R_{max}-R_{min})*(x+1)/2,
\label{mapno2}
\end{equation}
(iii) MAPNO=3 (Ref.~\cite{THK,TK}): 
\begin{equation}
y(x)=R_{min}+\frac{R_{d}{\rm tan}(\frac{\pi}{4}(1+x))}{1+\frac{R_{d}}{R_{max}
-R_{min}}{\rm tan}(\frac{\pi}{4}(1+x))},
\label{mapping}
\end{equation}
where
\begin{equation}
R_{d}=\frac{R_{med}-R_{min}}{R_{max}-R_{med}}(R_{max}-R_{min}).
\end{equation}
Because of its flexibility, we use the tangent mapping. According to the tangent 
mapping, 
\begin{equation}
y(-1)=R_{min},\,\,y(0)=R_{med},\,\,y(1)=R_{max}.
\end{equation}
Therefore, one can safely control the range $(R_{min},R_{max})$ and the 
distribution $(R_{med})$ of the grid points. With this discretization procedure, 
continuous 
integral equations are transformed into 
nonsingular matrix equations.
\section{{\bf Description of input parameters and input distribution}}
\label{INPUT}
\setcounter{equation}{0}
\setcounter{figure}{0}
\setcounter{table}{0}

\subsection{Input parameters}\label{PARAMET}

\vspace{0.3cm}

\begin{flushleft}
\small
\noindent
\begin{tabular}{|l|l|} 
\hline
 NF    & Number of quark flavors \\
\hline
 NGRID & Number of grid points  \\ 
\hline
 PLQCD  & $\Lambda_{qcd}$ \\ 
\hline
 Q2I      & Initial $Q^2$ \\ 
\hline
 Q2F      & Final $Q^2$ \\ 
\hline
 RMIN     & Smallest grid point \\ 
\hline
 RMED     &  Median of grid poiunts\\ 
\hline 
 RMAX     & The maximum grid point \\
\hline 
 LN       & The highest degree of Laguerre polynomial included \\
\hline
\end{tabular}

\begin{tabular}{|l|c|l|} \hline
 IFLAV & 1 & u quark  \\
\cline{2-3}
       & 2 & d quark\\
\cline{2-3}
       & 2 & s quark\\ 
\hline
MAPNO & 1 & type-1 mapping\\
\cline{2-3}
      & 2 & linear mapping\\
\cline{2-3}
      & 3 & tangent mapping\\
\hline
IALPH   & 0 & leading order (LO) in $\alpha_s$ \\ 
\cline{2-3}
        & 1 & next-to-leading order (NLO) in $\alpha_s$ \\ 
\hline
\end{tabular}
\end{flushleft}

\subsection{Arrays }\label{ARRAy}

\begin{flushleft}
\begin{tabular}{|l|l|} \hline
 P(I) & Grid points $y=P(I)$, where $RMIN<P(I)<RMAX$  \\
\hline
 WP(I) & Weights\\
\hline
 PT(I) & Grid points $t=PT(I)$, where $0<t<T=-2/\beta_0 \ln(\alpha_s(q^2)/\alpha_s(q^2_0))$  \\
\hline
 WT(I) & Weights\\
\hline
 PN(LN,IE,JE)  & elements of the LN'th Laguerre moment of LO kernel $[P]_{2\times2}$  \\ 
\hline
 RN(LN,IE,JE)  &  elements of the LN'th Laguerre moment of NLO kernel $[R]_{2\times2}$  \\ 
\hline
 SPN(I,IE,JE)  & PN(I,IE,JE)- PN(I-1,IE,JE) \\
\hline
 E1(J,K)       & PN(0,J,K)/$\lambda$   \\
\hline
 E2(J,K)       & -PN(0,J,K)/$\lambda$  \\
\hline
SA(K,N,IE,JE)  & $[a^k_n]_{2\times 2}$ \\
\hline
SB(K,N,IE,JE)  & $[b^k_n]_{2\times 2}$ \\
\hline
A(K,N,IE,JE)   & $[A^k_n]_{2\times 2}$ \\
\hline
B(K,N,IE,JE)   & $[B^k_n]_{2\times 2}$ \\
\hline
ENT(I,IE,JE)   & $[E_n(t)]_{2\times 2}$ \\
\hline
PHI0(Y,IE) & Initial distributions $x\Delta\Sigma_0$, $x\Delta G_0$ respectively for $IE=1,2$\\
\hline
PHI0N(N,IE)   & Laguerre moments of initial distribution. \\
\hline
PHIT(I,IE) & Evolved distributions $x\Delta\Sigma$, $x\Delta G$ respectively for $IE=1,2$\\
\hline
\end{tabular}
\end{flushleft}

The codes provided are stored in three directories named after the 
polarization types; namely transverse, longitudinal and unpolarized.
In each directory there are exacutable files that can be used to 
compile the fortran codes (executable file ``comp'') and run them (executable
file ``run''). The transverse polarization case is the simplest. For transverse polarization one has only the nonsinglet equation. In the longitudinally 
polarized and unpolarized cases both singlet and nonsinglet equations are 
solved.
The solutions of the singlet and nonsinglet equations are provided by separate codes. 
For the longitudinally polarized case the nonsinglet equation is 
solved by ``nslong.f'', while the singlet equation  is solved by ``snlong.f''. 
Both codes use the same input file, which is called ``INPUT''. Nonsinglet and 
singlet equations work in coordination with 
each other. Upon execution of the ``run'' command, first the nonsinglet 
equation is solved, and the output is written into data files. Next the 
nonsinglet equation is solved. The data produced by the solution of the singlet
equation is read by the code which solves the nonsinglet equation. The user is expected
to prepare the ``INPUT'' file, compile the codes by using command ``comp'' 
and then run them using the command ``run''. 
The procedure for the unpolarized case is the same, with the only difference being the names of the 
files which are ``nsunpol.f'' and ``snunpol.f''.  Next, we give the detailed
descriptions of the programs ``sn*.f''

\section{SNLONG and SNUNPOL}
Since the only difference between these programs are the kernels and the initial 
conditions, the explanations provided below equally apply to both.  

\subsection{Main Program}
The main program starts by calling the INPUT subroutine. This subroutine reads
12 input parameters described in Table~\ref{PARAMET}. 
Next, the grid points and the weights are calculated by calling the subroutine GAUS, and 
they are mapped into the desired interval. The mapping is done by the MAP subroutine.

Then, in order to calculate the Laguerre moments of the kernel and input 
distributions, the subroutine XPL is called. The factors $A^k_n$, $B^k_n$ which 
are used in the construction of the leading order evolution operator $E^0_n$ are 
calculated in a subroutine called XBKN. In the next step, the XEN subroutine 
is called to calculate the evolution operator $E_n$. Finally, the evolution 
of the initial distribution is performed in the VQD subroutine. The results
for the evolved distributions $\Delta \Sigma (x)$, and $\Delta G(x)$ are written,
respectively, into the files ``dltsig.dat'', and ``dltglu.dat''. Later, the result
for the $\Sigma$ distribution is read by the program that solves the 
nonsinglet equation.  
\subsection{Subroutine INPUT}
In this subroutine 12 input parameters are read. In addition to 
reading these parameters, various constants to be used throughout the program
are also defined in the INPUT subroutine. These constants are 
${\rm PLQCD}=\lambda_{QCD}$, ${\rm PI}=\pi$, ${\rm Z3}=\zeta_3$, CF, CG, CA, TR, ${\rm BT0}=\beta_0$,
${\rm BT1}=\beta_1$, $T=-2/\beta_0\ln(\alpha_s(q^2 )/\alpha_s(q^2_0))$, and 
${\rm XLMBD}=\lambda$. 
\subsection{Subroutine XPL}
In this subroutine we calculate the Laguerre moments of the initial distributions
PHI0(Y,IE) and kernels $P_0$ and $R=P_1-\beta_1/(2\beta_0)P_0$. In PHI0(Y,IE),
\begin{equation}
Y=\ln(1/X)\equiv P(J),
\end{equation}
 and the IE index is used to label the initial distributions for 
$x\Delta\Sigma(x)$(IE=1), and $x\Delta G(x)$(IE=2). We use variable $Y=[0,\inf)$ 
rather than $X=[0,1]$ in calculating the Laguerre moments.
The results for the Laguerre
moments of the initial distributions are stored in PHI0N(I,IE) where $IE=1,2$ 
corresponds to the Laguerre moments of $x\Delta\Sigma,x\Delta G$ respectively,
and $I$ refers to the order of Laguerre moments. 
In order to calculate the Laguerre moments of the kernel, we express it 
in the following form
\begin{equation}
P=P_r+\frac{P_s}{(1-x)_+}+P_\delta\delta(1-x).
\end{equation} 
The singular part $P_s$ is regulated according to the ``+'' regularization 
prescription, while the regular part $P_r$ is directly integrated. The delta 
function part requires no numerical integration. Therefore, the contribution 
of this piece is trivially added after performing the numerical integrals for $P_r$ and 
$P_s$. 
The kernels are stored 
in the arrays P0R(Y,IE,JE), P0S(Y,IE,JE), P1R(Y,IE,JE), P1S(Y,IE,JE), where 
``R'' and ``S'' stand for the ``regular'' and ``singular'' parts of the kernels, 
and $IE=1,2$, $JE=1,2$ are matrix indices. According to our convention, the distributions 
are introduced by define $2\times1$ vector array as $(x\Delta\Sigma,x\Delta G)$. 
In this subroutine
we also define the arrays SPN(I,IE,JE), E1(J,K), and E2(J,K).

\subsection{Subroutine XBKN}
In this subroutine we construct the arrays SA(K,N,IE,JE), SB(K,N,IE,JE),
A(K,N,IE,JE), B(K,N,IE,JE) using nested loops.

\subsection{Subroutine XEN}
This is the subroutine where the Laguerre moments of the evolution operator, 
ENT(N,IE,JE), are constructed. In this subroutine the functions E0N(N,T,IE,JE)
and E1N(N,T,IE,JE) are called. These functions represent the 0th and 1st order
contributions to the Laguerre moments.

\subsection{Function E0N(N,T,IE,JE)}
Calculates E0N(N,T,IE,JE) using the previously calculated A(K,N,IE,JE), and 
B(K,N,IE,JE) arrays. Notice that we have defined E0N(N,T,IE,JE) as a function
rather than a subroutine. The reason for this is the following: 
E0N(N,XT,IE,JE) appears inside an integral in the calculation of 
E1N(N,T,IE,JE), where XT($0<XT<T$) is the integration variable. Therefore, 
on needs to know E0N for all possible values of XT.   

\subsection{Function E1N(N,T,IE,JE)}
Calculates E1N. The set of grid points
for the XT integral is PT(I), and the weights are WT(I). Grid points, which was 
determined in the mained program using Gauss-Legendre subroutines, are chosen 
such that $0<PT(I)<T$. 

\subsection{Subroutine VQD}
Evaluates the final result PHIT(I,IE) for evolved distributions. Here, ``I'' 
represents the x (Bjorken) value, where $x=e^(-P(I))$, and IE as usual refers to two different
parton distributions, that is $PHIT(I,1)=xG(x)$, and $PHIT(I,2)=x\Sigma(x)$ (IE=2).

\subsection{Functions XLAG, FCTRL, S2, S1, ST}
XLAG(N,Y) computes the Laguerre polynomial of order N at Y. FCTRL(N) computes
N!. The result is given in double precision. S2(X) and S1(x) are respectively the 
functions $S_1(x)$, and $S_2(x)$. ST(x) represents $\tilde{S}(x)$.

\subsection{Other functions}
We also define various simple functions which are used throughout the program.
$ALPHAS(Q^2)$ is $\alpha_s(Q^2)$. GS(X,A,B,C,D,E,F) is the Gehrmann-Stirling ansatz ``A'' \cite{GS}
for the polarized parton distributions. In addition, for convenience in typing in the kernel for longitudinal
parton distributions, we define PF(X), PGS(X), PGR(X), PNFS(X),
PNFR(X), PA(X), FQQ(X), F1QG(X), F2QG(X), F1GQ(X), F2GQ(X), F3GQ(X), F1GGS(X),
F1GGR(X), F2GG(X), F3GGS(X), F3GGR(X).

\subsection{subroutines GAUSS, LEGENDRE and MAP}
The LEGEND(X,L,PSUBL) subroutine computes Legendre polynomials of argument 
x from 0 to order L. The GAUS(Y,WY,N) subroutine determines N gaussian points(in the vector Y) and N weights(in the vector WY). For this routine N must be even and not greater than 100. Since the points and weights are symmetric about 
zero, only half are stored. The original version of this subroutine was written by S. Cotanch \cite{SC}, at the Univ. of Pittsburgh in the years 1974-75.
The  MAP(Y,WY,N,MTYPE,RMIN,RMAX,RMED) subroutine maps the initial 
set of N grid points Y(I) and weights WY(I) to the desired interval (RMIN,RMAX).  
As explained in detail earlier, the mapping type MTYPE=3 allows one to control
the median(RMED) of the distribution. For the Y variable that we use in calculating the Laguerre moments, we have used RMIN=0.0D0, RMAX=1.0D2, and RMED=5.0d0
Since the Laguerre moment integrals are damped by exponential factors $e^{-Y}$,
 RMAX=1.0D2 was a large enough cutoff for the integral. 

\section{NSLONG and NSUNPOL}
The structure of the codes NSLONG and NSUNPOL are the same as that of SNLONG 
and SNUNPOL. All subroutine names and their functions are identicle. The 
only major difference comes from the fact that in the nonsinglet case one 
no longer has a matrix equation. Rather, there are two uncoupled equations.
The nonsinglet codes read the $x\Sigma(x)$ results which are produced by the 
singlet codes. 

\section{Running the code}
In order to get reliable results for $0.001<x<1$ we have used approximately 
30 Laguerre polynomials. The number of grid points used was $N=60$. The 
nonsinglet codes run in about 1 minute. The singlet and nonsinglet codes 
together run in about 20 minutes. At very small x $(x < 0.001)$ the Laguerre 
polynomials diverge, and therefore they are not a convenient basis to use in 
the very small x region. However, for $0.001<x<1$, the results are stable for 
a reasonable number of Laguerre polynomials.

\section{Conclusions}

We have shown that the Laguerre expansion is a significant tool 
in the analysis of the QCD evolution equations from the numerical side. 
These advantages include both short running times in the actual implementation 
of the evolution and the possibility to have well defined 
recursion relations. The Laguerre algorithm is a very powerful 
way to address efficiently these problems. 

We remark that polynomial 
expansions are going to be of wide use in the analysis 
of more general parton distributions -such as the non-forward or the 
double distributions- which have been introduced in the recent 
literature on Compton processes. Here we have just began our tour 
on the analysis of QCD renormalization group equations and their 
solutions by finite step integrations. We hope to return 
to the study of the extension of these methods to the 
non-diagonal partonic evolution in the near future. 

\centerline{\bf Acknowledgements}
We are grateful to L.E. Gordon, Gordon Ramsey for discussions.

\begin{figure}[thb]
\begin{center}
\mbox{
   \epsfxsize=6.0in
   \epsfysize=4.5in
\epsfbox{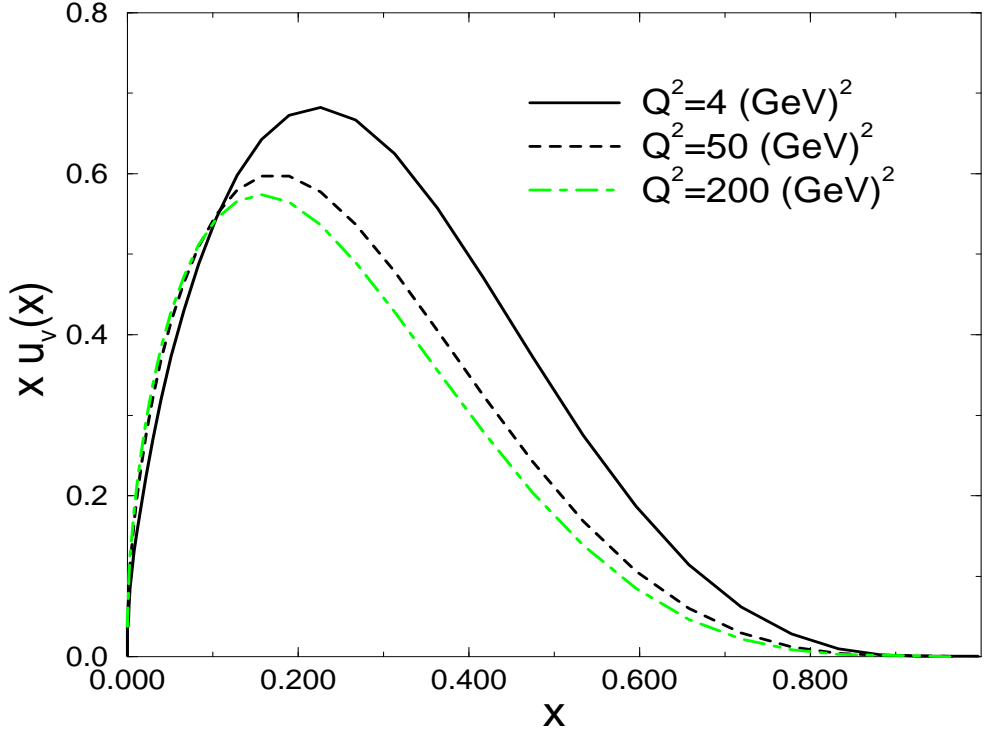}
}
\end{center}
\caption{Evolution of the 
unpolarized valence quark distributions ${x\rm u_v}$ versus x 
for various $Q^2$ values}
\label{uv_unpol}
\end{figure}

\begin{figure}[thb]
\begin{center}
\mbox{
   \epsfxsize=6.0in
   \epsfysize=4.5in
\epsfbox{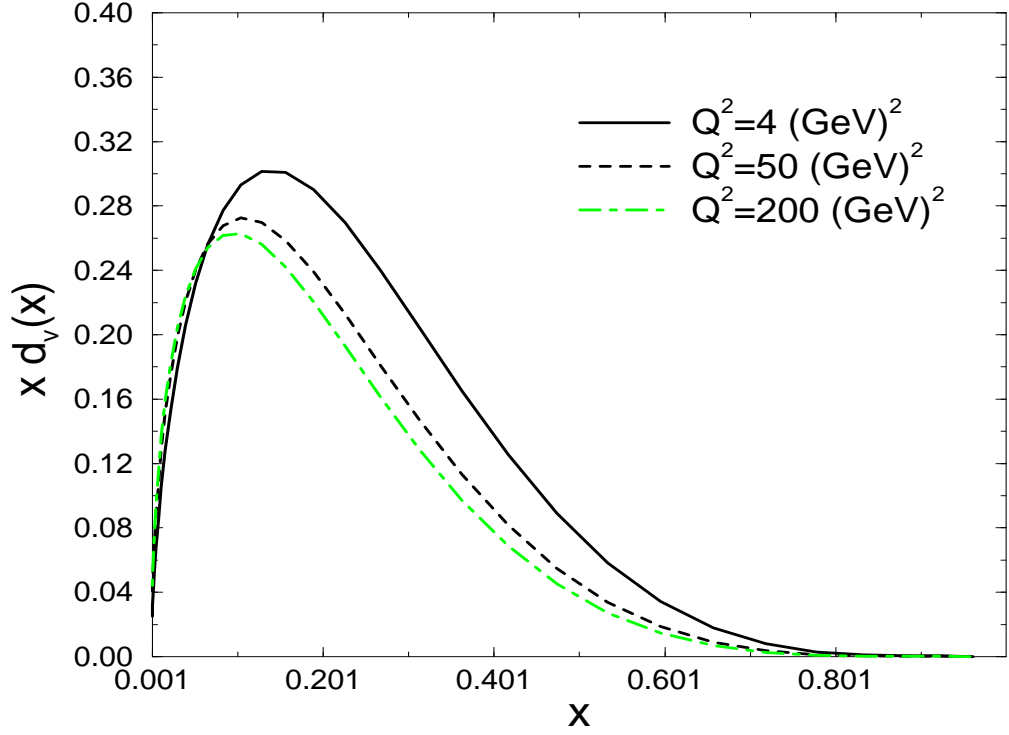}
}
\end{center}
\caption{Unpolarized ${x\rm d_v}$ versus x for various $Q^2$ values}
\label{dv_unpol}
\end{figure}

\begin{figure}[thb]
\begin{center}
\mbox{
   \epsfxsize=6.0in
   \epsfysize=4.5in
\epsfbox{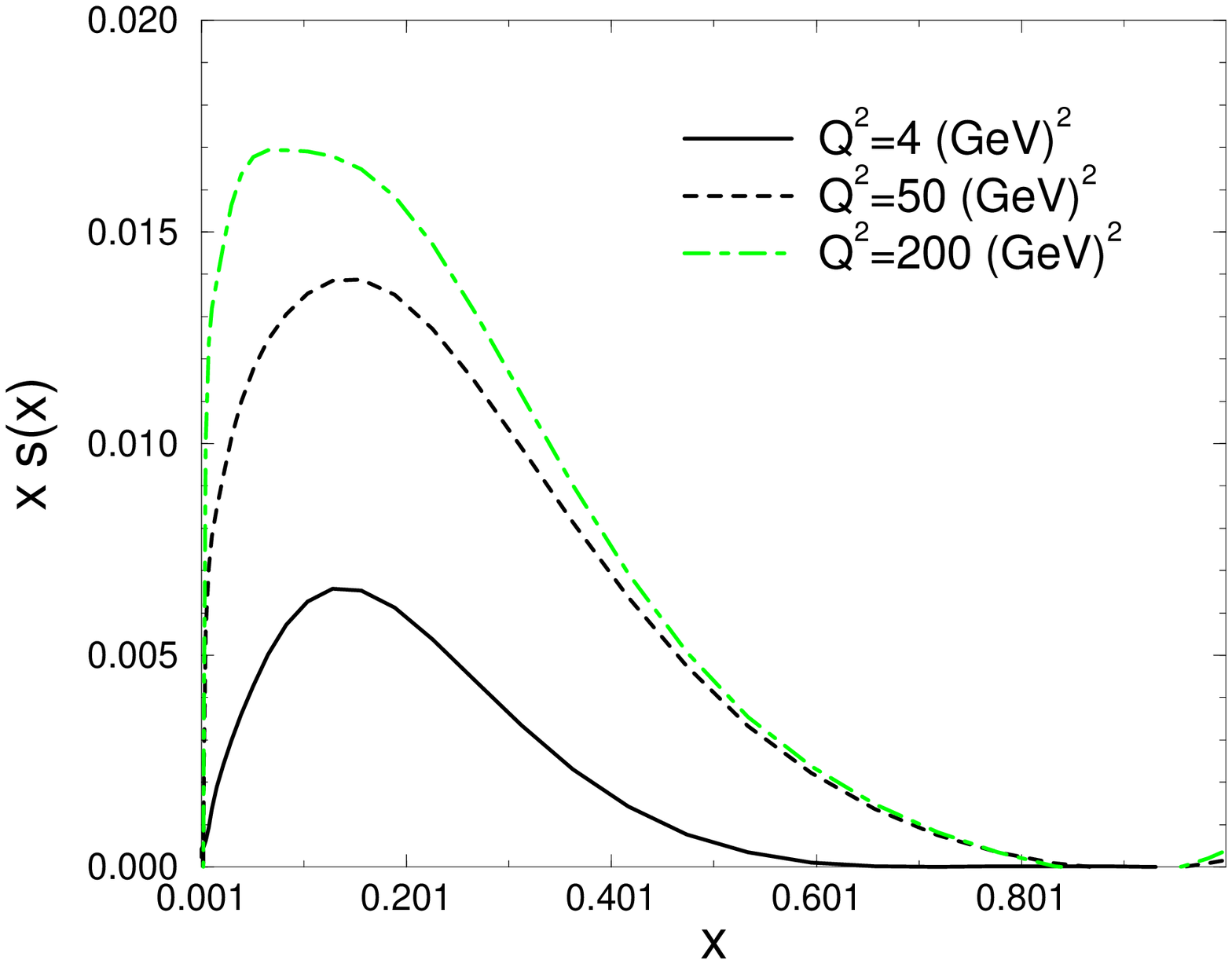}
}
\end{center}
\caption{Unpolarized ${x\rm s}$ distribution versus x for various $Q^2$ values}
\label{s_unpol}
\end{figure}

\begin{figure}[thb]
\begin{center}
\mbox{
   \epsfxsize=6.0in
   \epsfysize=4.5in
\epsfbox{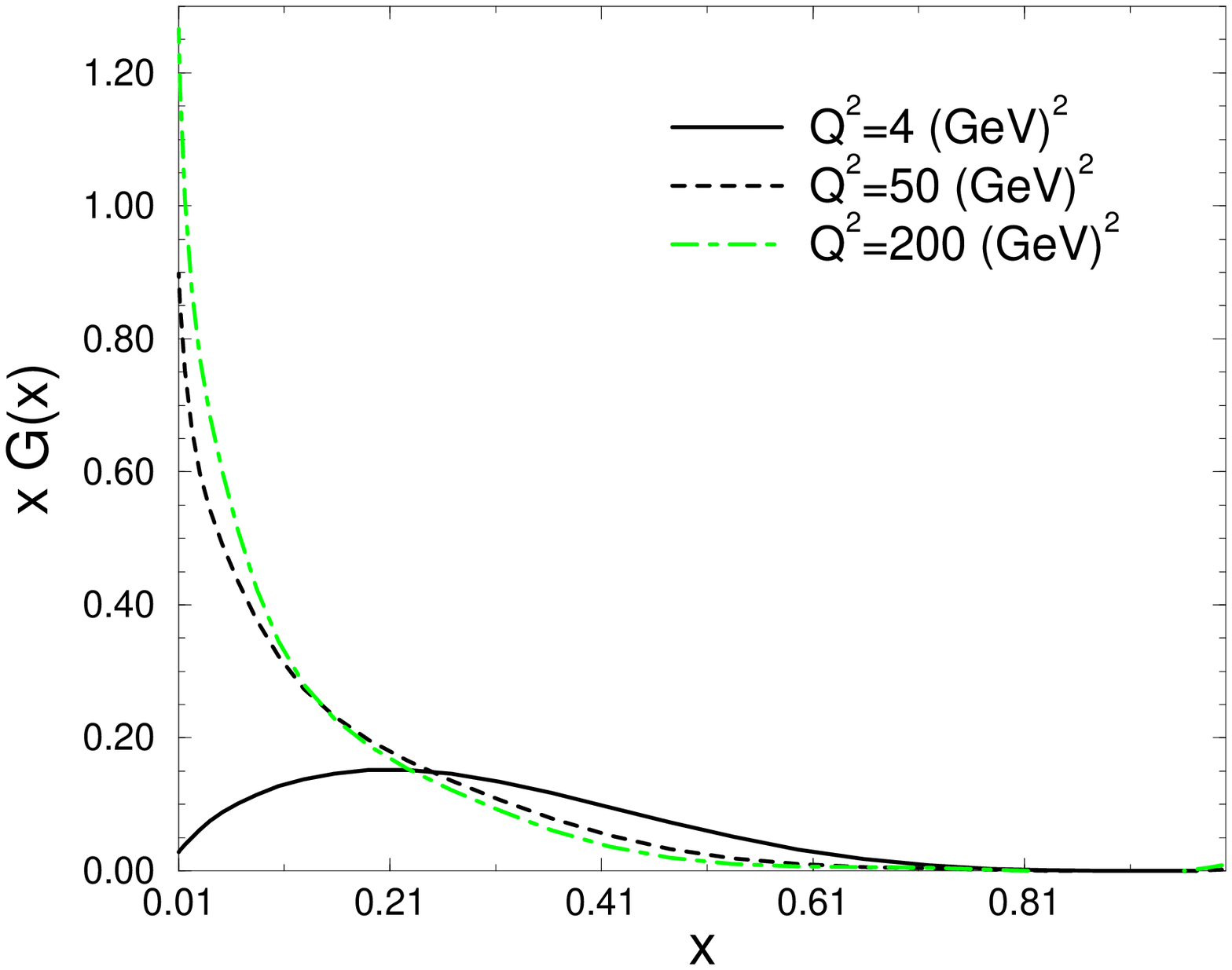}
}
\end{center}
\caption{Unpolarized gluon distribution 
${x\rm G}$ versus x for various $Q^2$ values}
\label{glu_unpol}
\end{figure}

\begin{figure}[thb]
\begin{center}
\mbox{
   \epsfxsize=6.0in
   \epsfysize=4.5in
\epsfbox{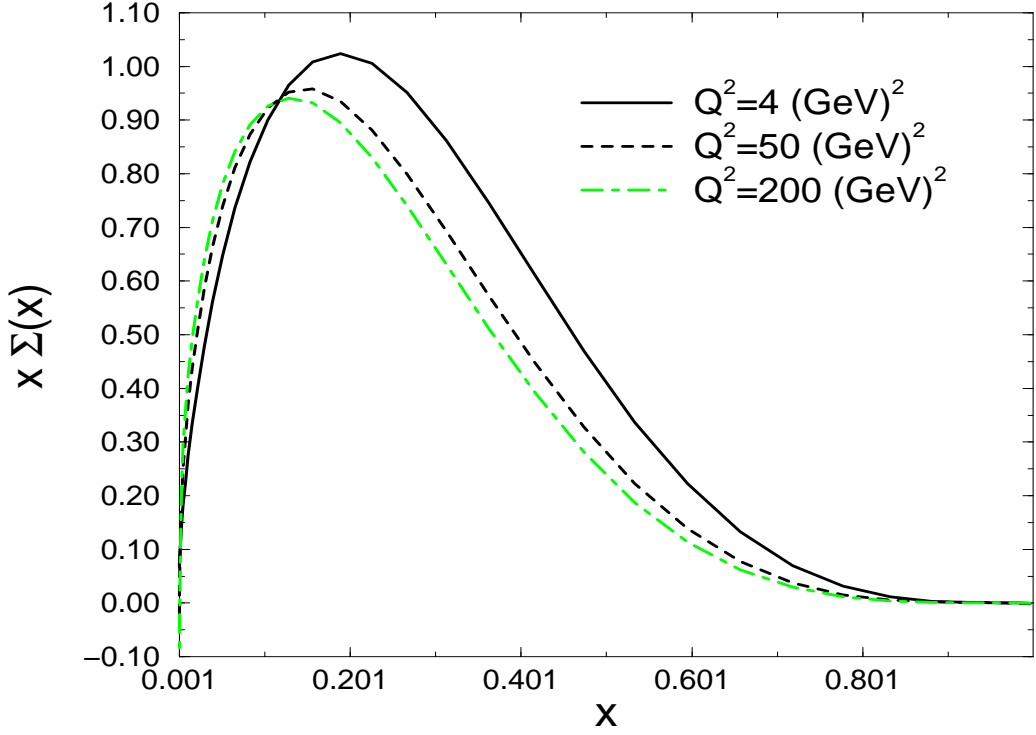}
}
\end{center}
\caption{Unpolarized $x\Sigma=x(u + d + s)$ versus x for various $Q^2$ values}
\label{sigma_unpol}
\end{figure}


\begin{figure}[thb]
\begin{center}
\mbox{
   \epsfxsize=6.0in
   \epsfysize=4.5in
\epsfbox{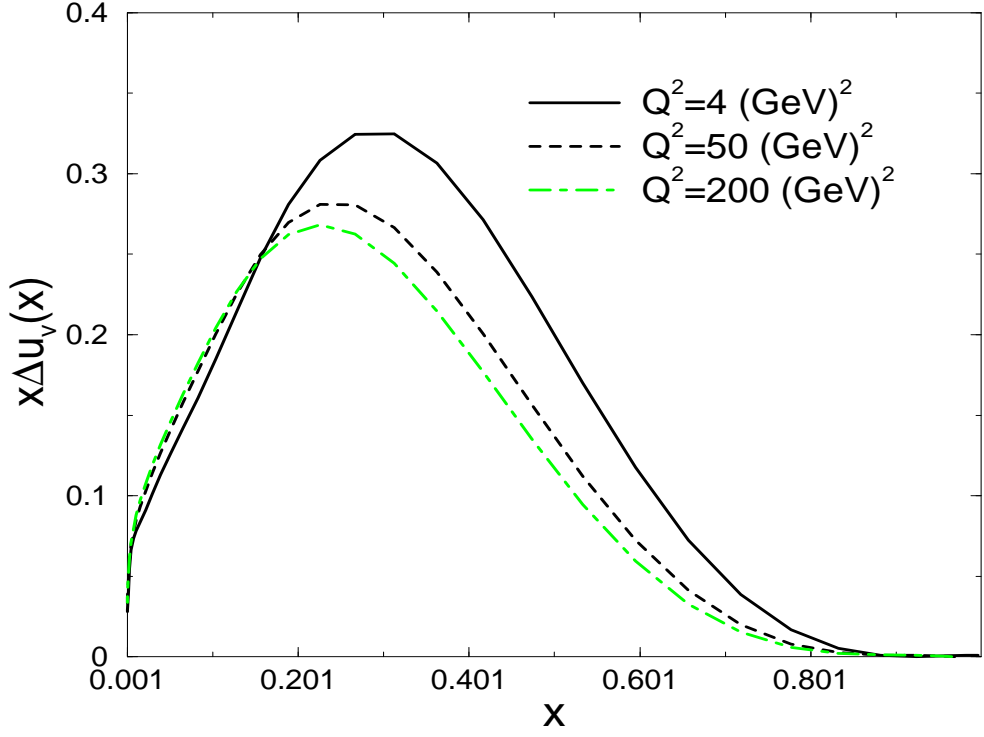}
}
\end{center}
\caption{Longitudinally polarized ${x\rm \Delta u_v}$, plotted versus x, 
 for various $Q^2$ values}
\label{uv_long}
\end{figure}

\begin{figure}[thb]
\begin{center}
\mbox{
   \epsfxsize=6.0in
   \epsfysize=4.5in
\epsfbox{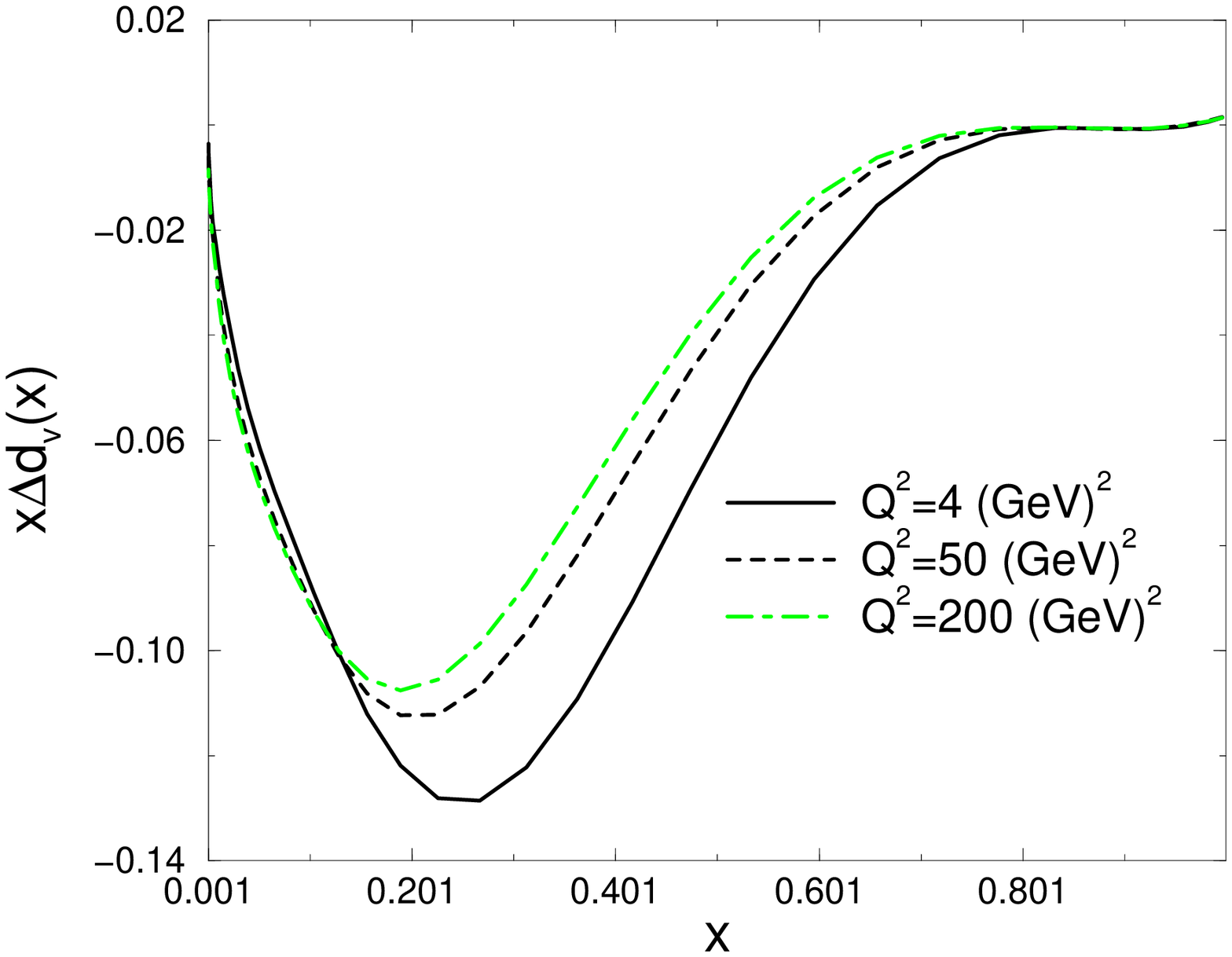}
}
\end{center}
\caption{Longitudinally polarized ${x\rm \Delta d_v}$ versus x for various $Q^2$ values}
\label{dv_long}
\end{figure}

\begin{figure}[thb]
\begin{center}
\mbox{
   \epsfxsize=6.0in
   \epsfysize=4.5in
\epsfbox{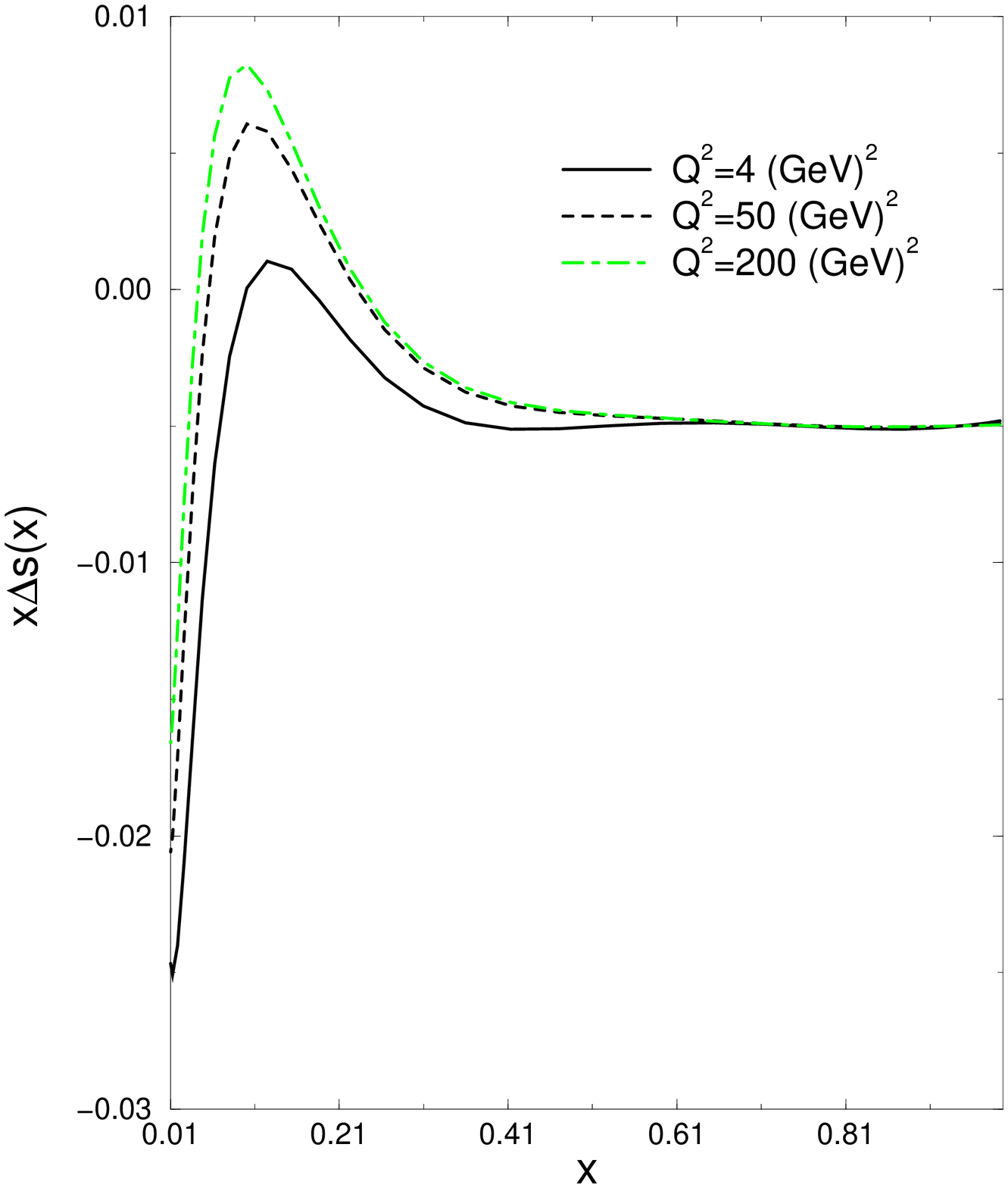}
}
\end{center}
\caption{Longitudinally polarized ${x\rm \Delta s}$ versus x for various $Q^2$ values}
\label{s_long}
\end{figure}

\begin{figure}[thb]
\begin{center}
\mbox{
   \epsfxsize=6.0in
   \epsfysize=4.5in
\epsfbox{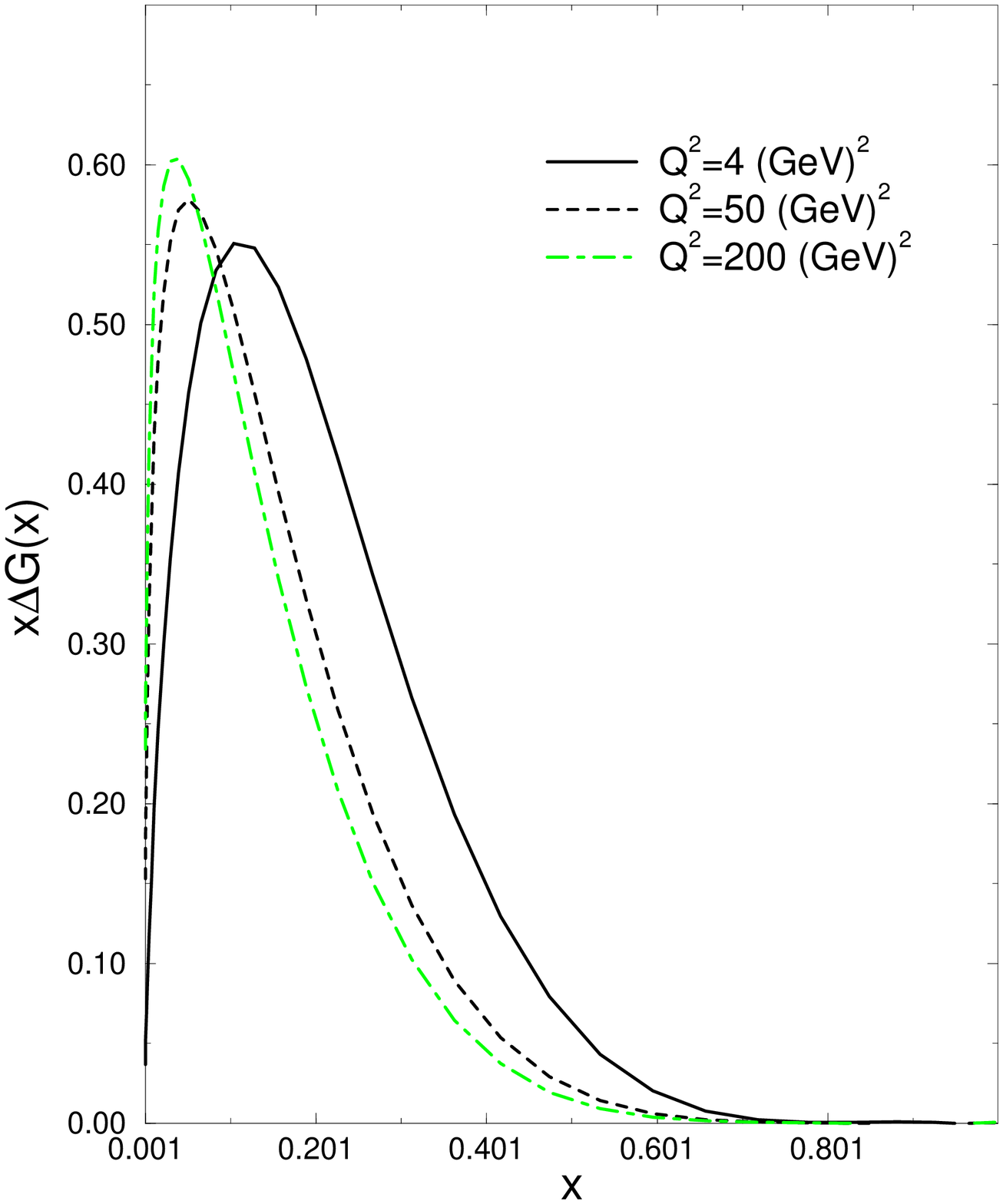}
}
\end{center}
\caption{Longitudinally polarized ${x\rm \Delta G}$ versus x, shown for various $Q^2$ values}
\label{glu_long}
\end{figure}

\begin{figure}[thb]
\begin{center}
\mbox{
   \epsfxsize=6.0in
   \epsfysize=4.5in
\epsfbox{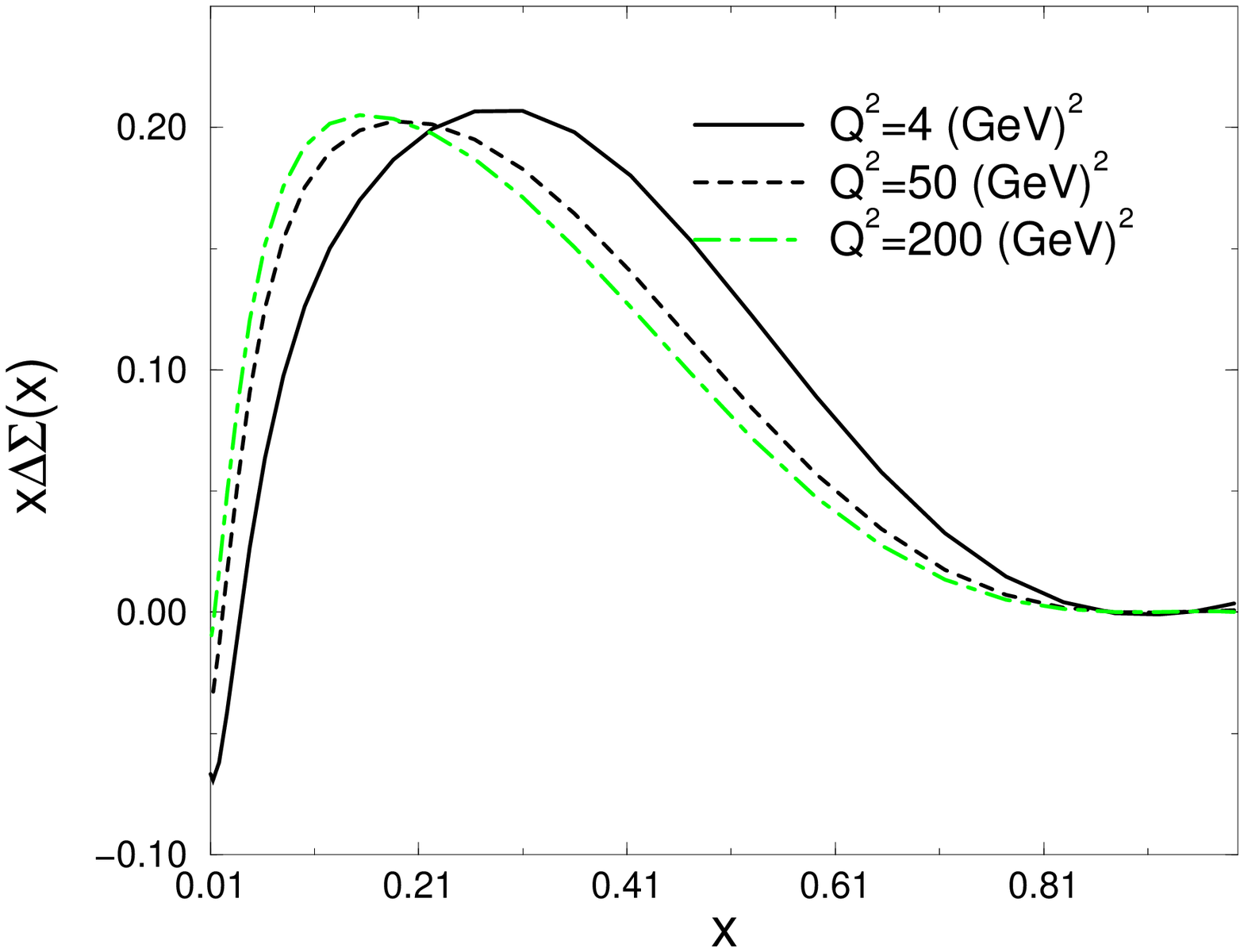}
}
\end{center}
\caption{Longitudinally polarized $x \Delta \Sigma$ versus x, shown for various $Q^2$ values}
\label{sigma_long}
\end{figure}


\begin{figure}[thb]
\begin{center}
\mbox{
   \epsfxsize=6.0in
   \epsfysize=4.5in
\epsfbox{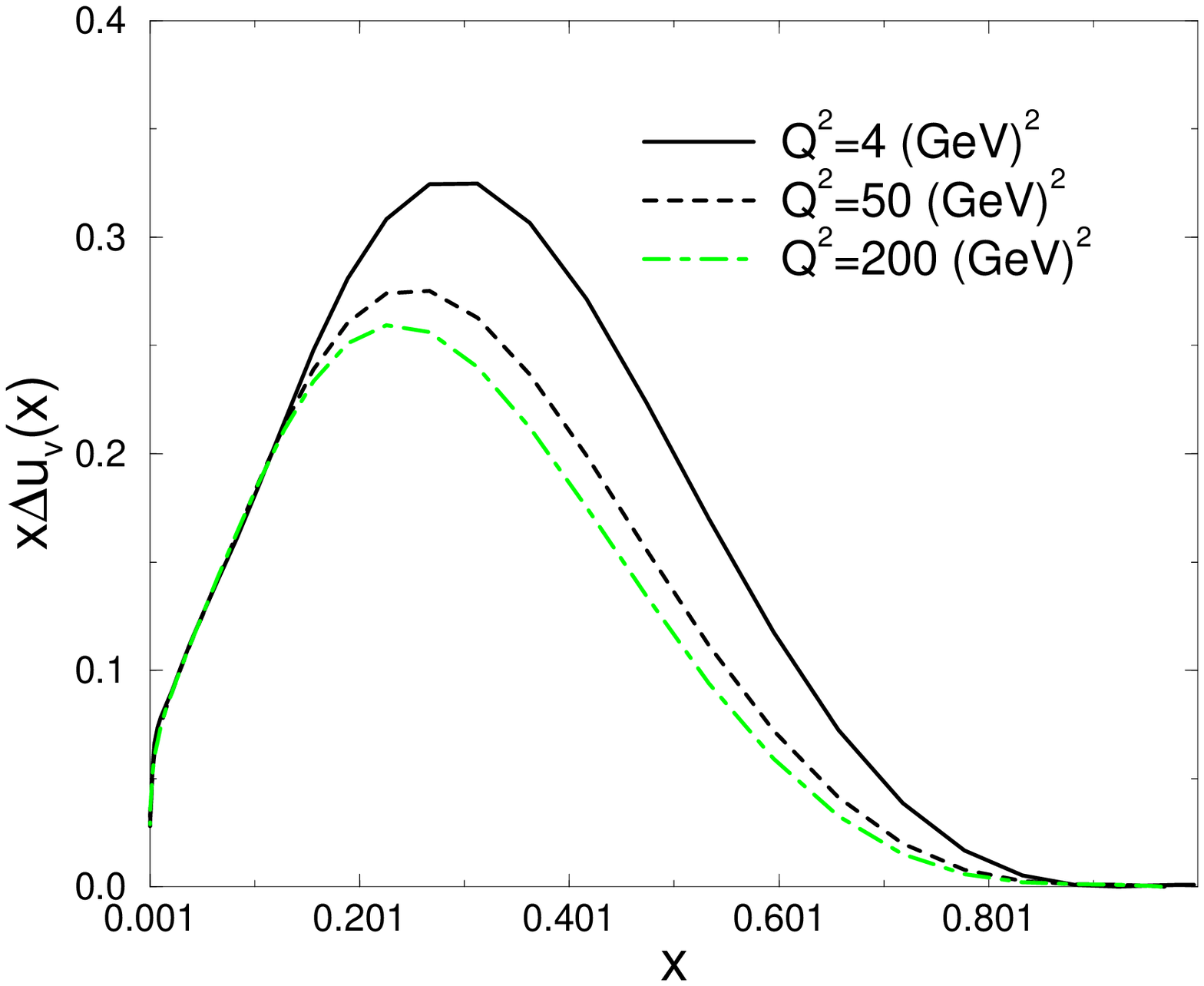}
}
\end{center}
\caption{Transversely polarized ${x\rm \Delta_T u_v}$ versus x, 
shown for various $Q^2$ values}
\label{uv_trans}
\end{figure}

\begin{figure}[thb]
\begin{center}
\mbox{
   \epsfxsize=6.0in
   \epsfysize=4.5in
\epsfbox{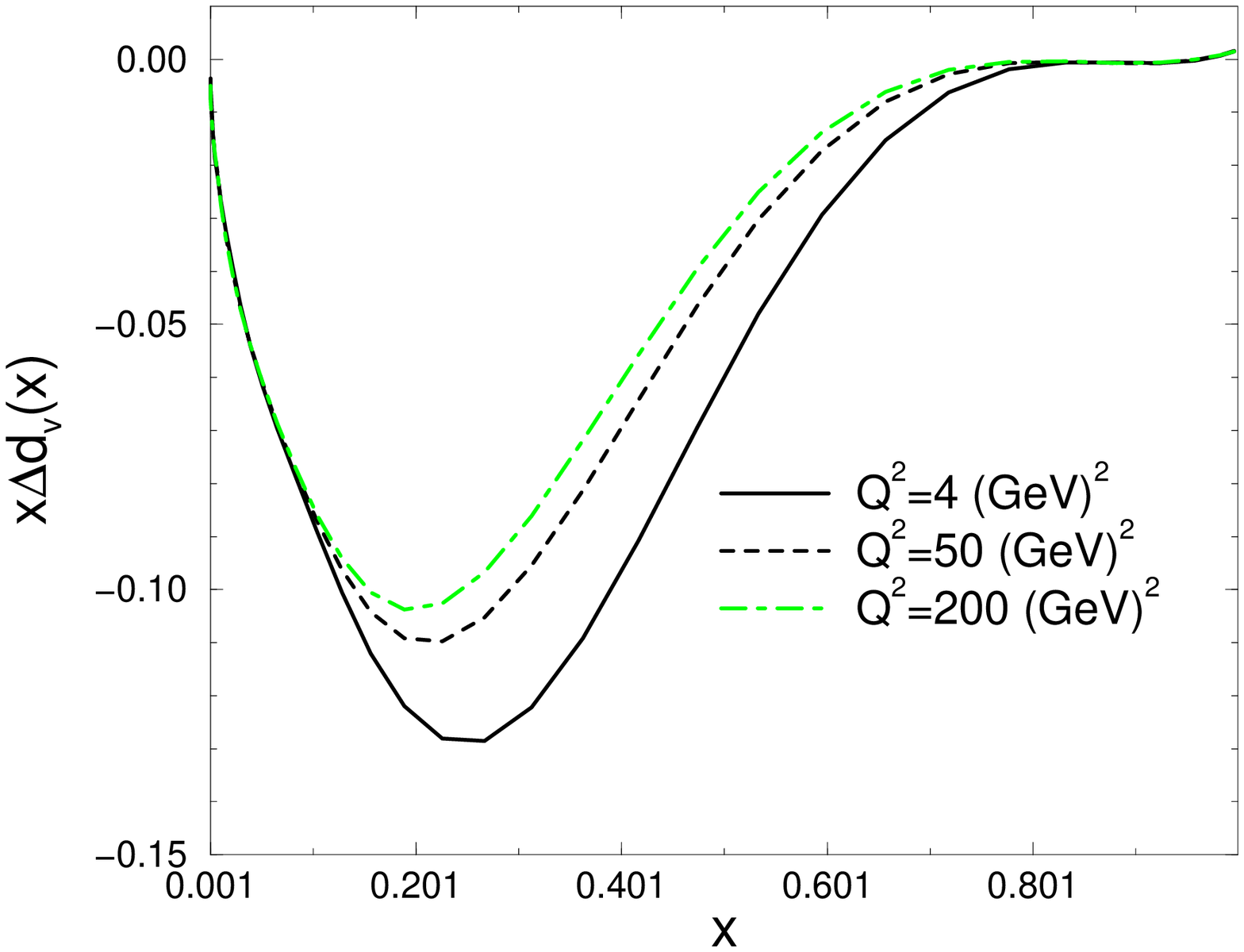}
}
\end{center}
\caption{Transversely polarized ${x\Delta_T d_v}$ versus x, 
shown for various $Q^2$ values}
\label{dv_trans}
\end{figure}

\begin{figure}[thb]
\begin{center}
\mbox{
   \epsfxsize=6.0in
   \epsfysize=4.5in
\epsfbox{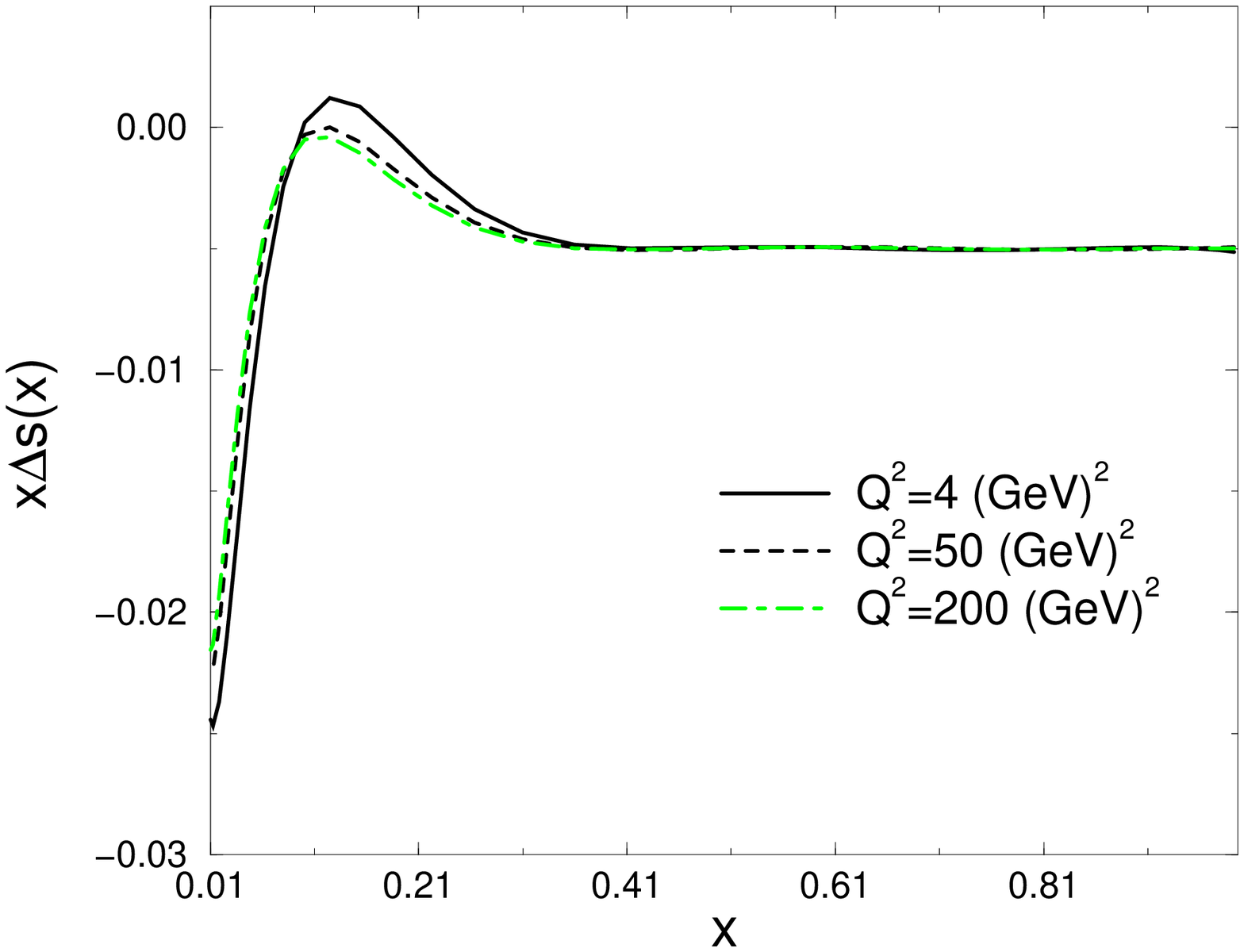}
}
\end{center}
\caption{Transversely polarized ${\rm x\Delta_T s}$, versus x, shown for various $Q^2$ values}
\label{s_trans}
\end{figure}

\section{Appendix. List of all the NLO polarized kernels}
The references for the unpolarized NLO kernels are \cite{FP1,FP2}. 
The longitudinally polarized kernels can be found either 
as anomalous dimensions or splitting functions in \cite{MVN, Vogel}
The Perturbative expansion of the kernels is 

\beqn
\Delta P_{ij}(x,\alpha_s)= 
\left( {\alpha_s\over 2 \pi}\right)\Delta P^{(0)}_{ij}(x) + 
\left( {\alpha_s\over 2 \pi}\right)\Delta P^{(1)}_{ij}(x)
+ ...
\eeqn
where $ij$ are flavour indices.  
The non-singlet and the singlet polarized LO kernels, 
respectively, are given by 

\beqn
&&\Delta P^{(0)}_{NS}(x)=P^{(0)}_{qq}(x)= C_F\left( {2\over (1-x)_+} -1-x + 
{3\over 2}\delta(1-x)\right) \nonumber \\
&&\Delta P^{(0)}_{qq}(x)=\Delta P^{(0)}_{NS}(x) \nonumber  \\
&&\Delta P^{(0)}_{qg}(x)=2 n_f T_R(2 x -1)\nonumber \\
&&\Delta P^{(0)}_{gq}(x)= C_F(2-x) \nonumber \\
&& \Delta P^{(0)}_{gg}(x)= 2 C_G\left( {1\over (1-x)_+} -2 x + 1\right) + 
{\beta_0\over 2}\delta(1-x).
\eeqn

The ``+'' distributions are defined  by 
\beq
\int_0^1 dx {f(x)\over (1-x)_+}=\int_0^1 dx {f(x)-f(1)\over 1-x}.
\eeq

The non-singlet NLO kernels are given by 

\beqa
&& \Delta P^{(1)}_{NS \pm}=C_F^2\left[P_F(x)\mp P_A(x) + 
\delta(1-x)\left({3\over 8}- {\pi^2\over 2} + 6 \zeta(3)\right)\right] 
\nonumber \\
&&\hspace{2cm} + {1\over 2}C_F C_A\left[P_G(x) \pm P_A(x) + 
\delta(1-x)\left({17\over 12} + {11\over 9}\pi^2 - 6 \zeta(3)\right)\right]
\nonumber \\
&&\hspace{2cm} + C_F T_R n_f \left[P_{N_F}(x) -\delta(1-x) \left( {1\over 6} 
+ {2\over 9}\pi^2\right)\right]
\nonumber \\
\eeqn

\beqn
&& P_F(x)= -2 {1 + x^2 \over 1-x} \ln x \ln(1-x) -
\left( {3\over 1-x} + 2 x\right)\ln x -{1\over 2}(1+x) \ln^2 x- 5 (1-x)
\nonumber \\
&& P_A(x)=2 \left({1+x\over 1 + x^2}\right)S_2(x) + 2 (1 + x)\ln x + 
4(1-x) \nonumber \\
&& P_G(x)={1+ x^2\over (1-x)_+}\left[ \ln^2 x + {11\over 3}\ln x + 
{67\over 9} -{\pi^2\over 3}\right] + 2(1 +x)\ln x + {40\over 3}(1-x)
\nonumber \\
&& P_{N_F}(x)={2\over 3}\left[{1+ x^2\over (1-x)_+}
\left(-\ln x - {5\over 3}\right) -2 (1-x)\right] \nonumber \\
\eeqn

where
\beq
S_2(x)=-2 {\it Li}_2(-x) - 2 \ln x \ln(1+x) +{1\over 2}\ln^2 x - {\pi^2\over 6}.
\eeq

The NLO polarized singlet kernels are given by 

\beqn
&& \Delta P^{(1)}_{qq}(x)= \Delta P^{(1)}_{NS +} + 
2 C_F T_R n_f \Delta F_{qq}\nonumber \\
&& \Delta P^{(1)}_{q g}(x)= C_F n_f T_R \Delta F^{(1)}_{qg}(x) + 
C_G n_f T_R \Delta F^{(2)}_{qg}(x)\nonumber \\
&& \Delta P^{(1)}_{gq}(x)=C_F n_f T_R \Delta F^{(1)}_{gq}(x) + 
C_F^2 F^{(2)}_{gq}(x) + C_F C_G \Delta F^{(3)}_{gq}(x)
\nonumber \\
&& \Delta P^{(1)}_{gg}(x)=- C_G T_R n_f \Delta F^{(1)}_{gg}(x) - 
C_F T_R n_f \Delta F^{(2)}_{gg}(x) + C_G^2 \Delta F^{(3)}_{gg}(x)
\eeqn
where $C_G=C_A= N_c$ and 

\beqn
&& \Delta F_{qq}(x)= 1-x - (1- 3 x)\ln x - (1 +x) \ln^2 x 
\nonumber \\
&& \Delta F^{(1)}_{qg}(x)= -22 + 27 x - 9 \ln x + 8(1-x) \ln(1-x) \nonumber \\
&&\hspace{2cm} + \delta p_{qg}(x)\left[ 2 \ln^2 x (1-x) - 4 \ln(1-x)\ln x + \ln^2 x 
- {2\over 3} \pi^2\right] \nonumber \\
&& \Delta F^{(2)}_{qg}(x)=2 (12- 11 x) -8(1-x)\ln (1-x) + 2(1 + 8 x)\ln x 
\nonumber \\
&& \hspace{2cm} -2 \left[ \ln^2 (1-x) - {\pi^2\over 6}\right]\delta p_{qg}(x) - 
\left[ 2 S_2(x) - 3 \ln^2 x \right] \delta p_{qg}(-x)
\nonumber \\
&& \Delta F^{(1)}_{gq}(x)=-{4\over 9}(x + 4) - {4\over 3}\delta p_{gq}(x)\ln (1-x)
\nonumber \\
&& \Delta F^{(2)}_{gq}(x)=-{1\over 2} -{1\over 2}(4 -x)\ln x - 
\delta p_{gq}(-x)\ln(1-x) +\left[ -4 - \ln^2(1-x) +{1\over 2}\ln^2 x \right]
\delta p_{qg}(x)
\nonumber \\
&& \Delta F^{(3)}_{gq}(x)=(4 - 13 x)\ln x +{1\over 3}(10 + x)\ln(1-x) +
{1\over 9}(41 + 35 x) +{1\over 2}\left[-2 S_2(x) + 3 \ln^2 x\right]\delta p_{gq}(-x)
\nonumber \\
&& \hspace{2cm} + \left[ \ln^2 (1-x) -2 \ln(1-x)\ln x - {\pi^2\over 6}\right]
\delta p_{gq}(x)
\nonumber \\
&& \Delta F^{(1)}_{gg}(x)=4 (1-x) +{4\over 3} (1+x)\ln x +
{20\over 9}\delta p_{gg}(x) +{4\over 3}\delta(1-x)
\nonumber \\
&& \Delta F^{(2)}_{gg}= 10(1-x) +2 (5-x)\ln x +2(1+x) \ln^2 x + \delta(1-x) \nonumber \\
&& \Delta F^{(3)}_{gg}(x)={1\over 3}(29-67 x)\ln x -{19\over 2}(1-x) + 
4(1+x)\ln^2 x -2 S_2(x)\delta p_{gg}(-x) \nonumber \\
&& \hspace{2cm} + \left[ {67\over 9} -4 \ln(1-x)\ln x + \ln^2 x -{\pi^2\over 3}\right]
\delta p_{gg}(x) +\left[ 3\zeta(3) +{8\over 3}\right]\delta(1-x),
\eeqn
with

\beqn
&& \delta p_{qg}(x)=2 x -1 \nonumber \\
&& \delta p_{g q}(x)= 2 -x \nonumber \\
&& \delta p_{gg}(x)={1\over (1-x)_+} - 2 x + 1.
\eeqn

The unpolarized kernels, to LO are given by 

\beq
P^{(0)}_{qq, NS}= \Delta P^{(0)}_{NS}
\eeq

for the non-singlet sector, and by 

\beqn
&& P^{(0)}_{qq}(x)=P^{(0)}_{qq,NS}\nonumber \\
&& P^{(0)}_{qg}(x)= 2 T_R n_f \left( x^2 + (1-x)^2\right)\nonumber \\
&& P^{(0)}_{gq}(x)=C_F {1 + (1-x)^2\over x}\nonumber \\
&& P^{(0)}_{gg}(x)= 2 N_c\left( {1\over (1-x)_+} + {1\over x}
-2 + x (1-x)\right) + {\beta_{o}\over 2}\delta(1-x)
\eeqn

in the singlet sector.

The NLO unpolarized non-singlet and singlet kernels are given by 
\beqa
P^{(1)\pm}_{NS}(x)=\Delta P^{(1)\mp}_{NS}(x)
\eeqa

and 
\beqn
&& P^{(1)}_{qq}(x)=  P^{(1)}_{NS +} + 2 C_F T_R n_f F_{qq}
\nonumber \\
&&  P^{(1)}_{q g}(x)= C_F n_f T_R  F^{(1)}_{qg}(x) + 
C_G n_f T_R F^{(2)}_{qg}(x)
\nonumber \\
&&  P^{(1)}_{gq}(x)=C_F n_f T_R  F^{(1)}_{gq}(x) + 
C_F^2 F^{(2)}_{gq}(x) + C_F C_G  F^{(3)}_{gq}(x)
\nonumber \\
&&  P^{(1)}_{gg}(x)=C_F T_R n_f  F^{(1)}_{gg}(x)
+C_G T_R n_f  F^{(2)}_{gg}(x) + 
 + C_G^2  F^{(3)}_{gg}(x)
\nonumber \\
\eeqn

\beqn
&& F_{qq}(x)={20\over 9 x} -2 + 6 x - {56\over 9} x^2 +
(1 + 5 x + {8\over 3} x^2 ) \ln x - (1 + x) \ln^2 x \nonumber \\
&& F^{(1)}_{qg}= 4 - 9 x -(1 - 4 x) \ln x -(1 - 2 x) \ln^2 x + 
4 \ln (1-x) \nonumber \\
&& + \left[ 2 \ln^2 \left({1-x\over x}\right) - 
4\ln\left({1-x\over x}\right) - {2\over 3} \pi^2 + 10
\right] p_{qg}(x) \nonumber \\
&& F^{(2)}_{qg}= {182\over 9} +{14\over 9}x + {40\over 9 x} +
({136\over 3}x - {38\over 3})\ln x - 
4\ln(1-x) -(2 + 8 x)\ln^2 x \nonumber \\
&& + \left[ -\ln^2 x +{44\over 3}\ln x - 2 \ln^2(1-x) + 4 \ln(1-x) +
{\pi^2\over 3} -{218\over 9}\right]p_{qg}(x) \nonumber \\
&& \hspace{1cm}+ 2 p_{qg}(-x) S_2(x)\nonumber \\
&& F^{(1)}_{gq}=-{4\over 3}x -\left[{20\over 9}+{4\over 3}\ln(1-x)
\right]p_{gq}(x)\nonumber \\
&& F^{(2)}_{gq}(x)=-{5\over 2}- {7\over 2}x +(2 +{7\over 2}x)\ln x -
(1-{1\over 2}x)\ln^2 x -2 x \ln(1-x)\nonumber \\
&& -\left[3 \ln(1-x) +\ln^2(1-x)\right]p_{gq}(x) \nonumber \\
&& F^{(3)}_{gq}(x)=\left({28\over 9}+{65\over 18}x +{44\over 9}x^2 -
(12 + 5 x +{8\over 3}x^2\right)\ln x +(4 + x)\ln^2 x + 2 x \ln(1-x)\nonumber \\
&& \hspace{1cm} + \left[-2 \ln x \ln(1-x) +{1\over 2}\ln^2 x +
{11\over 3}\ln(1-x) -{\pi^2\over 6} +{1\over 2}+ \ln^2(1-x)
\right]p_{gq}(x)\nonumber \\
&& \hspace{1cm} +S_2(x)p_{gq}(-x)   \nonumber \\
&& F^{(1)}_{gg}(x)= -16 + 8 x +{20\over 3}x^2 +{4\over 3 x} -
(6 + 10 x)\ln x -2(1+x)\ln^2 x -\delta(1-x)\nonumber \\
&& F^{(2)}_{gg}(x)=2 - 2 x +{26\over 9}(x^2 -{1\over x}) 
-{4\over 3}(1+x)\ln x -{20\over 9}p_{gg}(x)-{4\over 3}\delta(1-x) \nonumber \\
&& F^{(3)}_{gg}(x)={27\over 2}(1-x) +{67\over 9}(x^2 -{1\over x})-
({25\over 3}-{11\over 3}x +{44\over 3}x^2)\ln x +4(1+x)\ln^2 x 
\nonumber \\
&& \hspace{1cm} +\left[{67\over 9}- 4 \ln x \ln(1-x) +\ln^2 x -{\pi^2\over 3}
\right]p_{gg}(x) +2 p_{gg}(-x) S_2(x)\nonumber \\
&& \hspace{2cm} +\delta(1-x) \left({8\over 3} +3 \zeta(3)\right)
\eeqn

We have set
\beqn
&& p_{qq}(x)={2\over (1-x)_+} -1-x\nonumber \\
&& p_{qg}(x)x^2 +(1-x)^2\nonumber \\
&& p_{gq}(x)={1 +(1-x)^2\over x} \nonumber \\
&& p_{gg}(x)={1\over (1-x)_+} +{1\over x} -2 + x  (1-x)
\eeqn

The kernel for the transverse polarization is written as 

\beq
\Delta_T P_{q^{\pm}}=\Delta_T P^{(0)}_{qq}(x) +
{\alpha_s(Q^2)\over 2 \pi} \Delta_T P^{(1)}_{q^{\pm}}(x)
\eeq

with the LO expression \cite{artru}
\beq
\Delta_T P_{q^{\pm}}=C_F\left[{2 x\over (1-x)_+} 
+{3\over 2}\delta(1-x)\right].
\eeq

The NLO corrections are given by \cite{vogel1}

\beqn
&& \Delta_T P^{(1)}_{q^{(\pm)}}(x)=\Delta_TP^{(1)}_{qq}(x) \pm 
\Delta_T P^{(1)}_{q\bar{q}}(x),  \\
&&\Delta_T P^{(1)}_{qq}(x)=C_F^2\left[1-x -
\left({3\over 2}+ 2 \ln(1-x)\right) \ln x \Delta_T P^{(0)}_{qq}(x)\right.
\nonumber \\
&& \left. + \left({3\over 8} -{\pi^2\over 2} +6\zeta(3)\right)\delta(1-x)
\right] +{1\over 2}C_F C_G\left[-(1-x) +
\left({67\over 9}+{11\over 3}\ln x + \ln^2 x-{\pi^2\over 3}\right)
\delta_T P^{(0)}_{qq}(x)\right.\nonumber \\
&&\left. +\left({17\over 12}+{11\over 9}\pi^2 -6 \zeta(3)\right)\delta(1-x)
\right] +{2\over 3}C_F T_R n_f \left[\left(-\ln x -{5\over 3}\right)
\delta_T P^{(0)}_{qq}(x) -
\left({1\over 4}+{\pi^2\over 3}\right)\delta(1-x)
\right),\nonumber \\
\eeqn
\beqn
&& \Delta_T P^{(1)}_{q\bar{q}}(x)=C_F\left(C_F-{1\over 2}C_G\right)
\left(-(1-x) +2 S_2(x)\delta_T P^{(0)}(-x)\right).
\eeqn


\begin{thebibliography}{99}
\bibitem{comm} Notice that eqs.~(3.26) and (3.37) of \cite{FP1} should read as in our eq.~(\ref{solve}) and 
eq.~(\ref{solve1}) respectively.
\end{thebibliography}
\end{document}